\documentclass[12pt,runningheads,a4paper]{article}
\usepackage{amsmath}
\usepackage{graphicx}
\usepackage{amssymb}
\usepackage[arrow,matrix]{xy}

\begin{document}
\date{}
%%%%%%%%%%%%%%%%%%%%%%%%%%%%%%%%%%%%%%%%%%%%%%%%%%%%%%%%%%%%%%%%%%%%%%%%%%%%%%%%%%%%%%%%%%%%%%%%%%%%%%%%%%%%%%%%%%%%%%%%%%%%%%%%%%%%%%%
\title{On the energetic balance for the flow of an Oldroyd-B fluid
induced by a constantly accelerating plate}%To be sent to "Appl. Math. Comput."
\maketitle
\author\begin{center}{Corina Fetecau${}^{\,a,}$}\footnote{Corresponding author. Tel. +40 232263218.\\
E-mail address: fetecau@math.tuiasi.ro; cfetecau@yahoo.de }, C. Fetecau${}^{\,b}$, A. Mahmood${}^{\,c}$, E. Axinte${}^{\,d}$
\end{center}
\begin{center}\scriptsize{\textit{ {$^{a}$Department of Theoretical Mechanics, Technical University of Iasi, R-6600 Iasi, ROMANIA\\$^{b}$Department of Mathematics, Technical University of Iasi, R-6600 Iasi, ROMANIA\\$^{c}$Abdus Salam School of Mathematical Sciences, 68-B, New Muslim Town, Lahore, PAKISTAN\\$^{d}$Department of Machine Manufacturing, Technical University of Iasi, R-6600 Iasi, ROMANIA}}}\end{center} \textbf{Abstract}\\\\{\footnotesize{
\indent Exact and approximate expressions are established for dissipation, the power of the shear stress at the wall and the boundary layer thickness corresponding to the motion of an Oldroyd-B fluid induced by a constantly accelerating plate. The similar expressions for Maxwell, Newtonian and second grade fluids, performing the same motion, are obtained as limiting cases of our general results. The specific features of the four models are emphasized by means of the asymptotic approximations.
\section{Introduction}

\indent \indent In recent years, the interest for flows of Newtonian and non-Newtonian fluids has considerably increased, while energetic aspects regarding these motions are scarcely met in the literature. The first results of this kind seem to be those of B\"{u}hler and Zierep [1] concerning the Rayleigh-Stokes problem for Newtonian fluids. These results have been recently extended by Zierep and Fetecau [2, 3] and Fetecau et al [4] to second grade, Maxwell and Oldroyd-B fluids. In their works, the authors also determine the boundary layer thickness and use it to obtain a series solution for the velocity field. Similar results for the flow induced by a constantly accelerating plate in second grade and Maxwell fluids have been also obtained by Fetecau et al in [5, 6]. For this motion, unlike the previous flow, the series solutions that have been obtained are completely determined by means of the corresponding initial and boundary conditions.

The aim of this paper is to extend the results from [5] and [6] to Oldroyd-B fluids. More exactly, we intend to present a complete study of the energetic balance for the unsteady flow of such a fluid, driven by the transversal motion of a constantly accelerating plate. Of special interest are: the Power due to the shear stress at the wall, Dissipation and the Changing of the kinetic energy with time. The first term describes the energy input that is necessary to keep the medium running. A decisive question is whether these terms are larger or smaller than in Maxwell, Second grade and Newtonian fluids. In order to see this, besides their exact values, approximative expressions have been established for small dimensionless relaxation and retardation times. Such expressions have been also achieved for velocity, shear stress and boundary layer thickness. Finally, in contrast with the previous results concerning the flow due to a suddenly moved plate [2-4, 7] and in accordance with those obtained in [5] and [6], the series solutions for velocity and shear stress have been completely determined by governing equations and the corresponding initial and boundary conditions. In the special cases, when the relaxation or retardation time tends to zero, all results that have been obtained reduce to those for second grade, respectively, Maxwell fluids. If both times are going to zero, the results for Newtonian fluids are recovered.

\section{Governing equations}

\indent\indent Recently, Rajagopal and Srinivasa [8] have built and developed a systematic thermodynamic framework within which models for a variety of rate type viscoelastic fluids can be obtained. Among them, the Oldroyd-B fluids that take into account elastic and memory effects exhibited by most polymeric and biological liquids have been used quite widely in many applications and the results of simulation fit experimental data in a wide range [9]. These fluids can describe stress-relaxation, creep and the normal stress differences that develop during simple shear flows. They store energy like a linearized elastic solid, their dissipation however being due to two dissipative mechanisms that implies that they arise from a mixture of two viscous fluids. The Cauchy stress in an incompressible Oldroyd-B fluid, is given by [8-10]
\begin{equation}
\textbf{T} = -p\,\textbf{I} + \textbf{S}, \,\,\,\,\,\,
\textbf{S}+\lambda(\dot{\textbf{S}}
-\textbf{L}\textbf{S}-
\textbf{S}\textbf{L}^{T})=\mu [\textbf{A}+\lambda_r(\dot{\textbf{A}}
-\textbf{L}\textbf{A}-
\textbf{A}\textbf{L}^{T})],                                                    %1
\end{equation}
where $-p\textbf{I}$  denotes the indeterminate spherical stress, $\textbf{S}$  is the extra-stress tensor, $\textbf{L}$  is the velocity gradient, $\textbf{A}=\textbf{L}+\textbf{L}^{T}$ is the first Rivlin-Ericksen tensor, $\mu$  is the dynamic viscosity, $\lambda$ and $\lambda_r$ are the relaxation and retardation times and the superposed dots indicate the material time derivatives. This model includes, as a special cases, the Maxwell and linearly viscous fluid models. Let us consider an incompressible Oldroyd-B fluid at rest, occupying the space above an infinitely extended plate which is situated in the plane $y=0$  of a Cartesian coordinate system $x$, $y$ and $z$. At time $t=0^+$, the plate is subject to a motion of constant acceleration $A$ in the $x$-direction. Owing to the shear the fluid above the plate is gradually moved, its velocity being of the form
\begin{equation}
\textbf{v}=\textbf{v}(y,t)=u(y,t)i,               %2
\end{equation}
where $i$  denotes the unit vector along the $x$-coordinate direction. For this flow, the constraint of incompressibility is automatically satisfied. Assuming that the extra-stress $\textbf{S}$ depends only of $y$ and $t$ and having in mind the initial condition
\begin{equation}
\textbf{S}(y,t)=\textbf{0},\,\,\,\,\,\,\,\,\,\,y>0,               %3
\end{equation}
the fluid being at rest up to the moment $t=0$, 0ne obtains (see for instance [11])
\begin{equation*}
S_{xz}=S_{yz}=S_{yy}=S_{zz}=0,
\end{equation*}
As regards the shear stress $\tau(y,t)=S_{xy}(y,t)$  satisfies the partial differential equation
\begin{equation}
(1+\lambda\partial_t)\tau(y,t)=\mu(1+\lambda_r\partial_t)\partial_yu(y,t).                             %4
\end{equation}

In the absence of pressure gradient in the flow direction and neglecting body forces, the balance of linear momentum leads to the meaningful equation
\begin{equation}
\partial_y\tau(y,t)=\rho\,\partial_tu(y,t);\,\,\,\,\,\,\,\,\,y,t>0,                                      %5
\end{equation}
Where $\rho$ is the constant density of the fluid. Eliminating $\tau(y,t)$  between Eqs. (4) and (5), one attain to the governing equation
\begin{equation}
\lambda\partial^2_tu(y,t)+\partial_tu(y,t)=\nu(1+\lambda_r\partial_t)\partial^2_yu(y,t);\,\,\,\,y,t>0,             %6
\end{equation}
where $\nu=\mu/\rho$  is the kinematic viscosity of the fluid, $\rho$ being its constant density.

The exact solutions of the partial differential equations (4) and (6) with the initial and boundary conditions
\begin{equation}
u(y,0)=\partial_tu(y,0)=0,\,\,\,\,\tau(y,0)=0\,\,\,\,\mbox{for}\,\,\,\,y>0,              %7
\end{equation}
\begin{equation}
u(0,t)=At\,\,\,\,\mbox{for}\,\,\,\,t\geq0;\,\,\,\,
u(y,t),\,\partial_yu(y,t)\rightarrow0
\,\,\,\,\mbox{as}\,\,\,\,y\rightarrow\infty.                                             %8
\end{equation}
are given by (cf. with [11], Eqs. (16) and (20))
\begin{equation}
u(y,t)=At-\frac{2A}{\nu\pi}\int_0^\infty
\bigg[1-\frac{r_2r_3\exp(r_1t)-r_1r_4\exp(r_2t)}{r_2-r_1}\lambda\bigg]
\frac{\sin(y\xi)}{\xi^3}\,d\xi                                                          %9
\end{equation}
and
\begin{equation}
\tau(y,t)=-\frac{2\rho A}{\pi}\int_0^\infty
\bigg[1-\frac{r_4\exp(r_2t)-r_3\exp(r_1t)}{r_2-r_1}\bigg]
\frac{\cos(y\xi)}{\xi^2}\,d\xi,                                                 %10
\end{equation}
\begin{equation*}
r_{1,2}=\frac{-(1+\alpha\xi^2)\pm\sqrt{(1+\alpha\xi^2)^2-4\nu\lambda\xi^2}}{2\lambda},
\end{equation*}
\begin{equation*}
r_{3,4}=\frac{1-\alpha\xi^2\pm\sqrt{(1+\alpha\xi^2)^2-4\nu\lambda\xi^2}}{2\lambda},
\end{equation*}
and $\alpha=\nu\lambda_r$.

By letting $\lambda_r$ or $\lambda\rightarrow0$  into Eqs. (9) and (10), ), we attain to the similar solutions for Maxwell (see [11], Eqs. (25) and (26)) and second grade (see [12], Eq. (2.5) with $V(t)=At$  or [13], Eq. (6) for the velocity field only) fluids. If both  $\lambda_r$ and $\lambda\rightarrow0$, the solutions [11]
\begin{equation*}
u_N(y,t)=At-\frac{2A}{\nu\pi}\int_0^\infty\bigg(1-\mbox{e}^{-\nu\xi^2t}\bigg)
\frac{\sin(y\xi)}{\xi^3}\,d\xi,
\end{equation*}
\begin{equation}
\tau_N(y,t)=-\frac{2\rho A}{\pi}\int_0^\infty\bigg(1-\mbox{e}^{-\nu\xi^2t}\bigg)
\frac{\cos(y\xi)}{\xi^2}\,d\xi,                                                           %11
\end{equation}
corresponding to a Newtonian fluid. The last solutions can be also obtained from (9) and (10) by making  $\lambda=\lambda_r$. This is not a surprise, it is a simple consequence of the Joseph's remark ([14], \S2.2). Of course, $u_N(y,t)$  and $\tau_N(y,t)$   given by Eqs. (9) can also be written in the equivalent classical forms (cf. [65], Eqs. (37) and (38))
\begin{equation}
u_N(y,t)=4At\,\mbox{i}^2\mbox{Erfc}\bigg(\frac{y}{2\sqrt{\nu t}}\bigg)
\,\,\,\,\mbox{and}\,\,\,\,\tau_N(y,t)=
-2\rho A\sqrt{\nu t}\,\,\mbox{iErfc}\bigg(\frac{y}{2\sqrt{\nu t}}\bigg),             %12
\end{equation}
where $\mbox{i}^n\mbox{Erfc}(x)=\int_x^\infty\mbox{i}^{n-1}\mbox{Erfc}(\xi)\,d\xi$  are the integrals of the complementary error function $\mbox{Erfc}(\cdot)$ and $\mbox{i}^0\mbox{Erfc}(x)=\mbox{Erfc}(x)$.

The energetic balance for a given volume $V$, as it results from [1-6], is given by
\begin{equation}
\frac{d}{dt}E_{kin}+L+\Phi=0,                                                   %13
\end{equation}
where $E_{kin}$ is the kinetic energy, $\Phi$  is the dissipation and $L$ is the power of the shear stress at the wall. The changing of the kinetic energy with time is given by
\begin{equation}
\frac{d}{dt}E_{kin}=\frac{d}{dt}\int_V\frac{\rho}{2}\textbf{v}^2dV=
\int_V\frac{\partial}{\partial t}
\bigg(\frac{\rho}{2}\textbf{v}^2\bigg)dV+
\int_A\frac{\rho}{2}\textbf{v}^2(\textbf{v.n})dA\,,                         %14
\end{equation}
where $A$ is the boundary of the flow domain and $\textbf{n}$ is the unit vector normal to $A$. For an infinite volume of rectangular cross-section with $x\in[0,l]$ and $z\in[0,1]$, Eqs. (2), (5), (14) imply
\begin{equation*}
\frac{d}{dt}E_{kin}=\rho\int_Vu(y,t)\frac{\partial u(y,t)}{\partial t}dV=\rho l\int_0^\infty u(y,t)
\frac{\partial u(y,t)}{\partial t}dy=l\int_0^\infty u(y,t)
\frac{\partial \tau(y,t)}{\partial t}dy=
\end{equation*}
\begin{equation}
-lu(0,t)\tau(0,t)-l\int_0^\infty \tau(y,t)\frac{\partial u(y,t)}{\partial y}dy.                        %15
\end{equation}
Comparing Eqs. (13) and (15), it results that
\begin{equation}
L=L(t)=lu(0,t)\tau(0,t)=lu_w(t)\tau_w(t)\,\,\,\,\mbox{and}\,\,\,\,\Phi=\Phi(t)=l
\int_0^\infty \tau(y,t)\frac{\partial u(y,t)}{\partial y}dy.                                          %16
\end{equation}

The boundary layer thickness, as it results from [7], is given by
\begin{equation}
\delta=\delta(t)=\frac{1}{u(0,t)}\int_0^\infty u(y,t)dy                          %17
\end{equation}
and represents the thickness of the fluid layer moved with the plate by friction. One measure of the boundary layer thickness is the distance from the wall where the velocity differs by 1 percent from the external velocity.

\section{Exact expressions for $L$, $\Phi$ and $\delta$}
Introducing the general solutions $u(y,t)$ and $\tau(y,t)$, given by Eqs. (9) and (10),
 into (16) and (17) we find for $L$, $\Phi$ and $\delta$ the exact expressions
\begin{equation}
L=-\frac{2\rho l A^2t}{\pi}\int_0^\infty
\bigg[1-\frac{r_4\exp(r_2t)-r_3\exp(r_1t)}{r_2-r_1}\bigg]
\frac{d\xi}{\xi^2},                                                    %1.3.1
\end{equation}
\begin{equation*}
\Phi=\frac{4\rho l A^2}{\nu\pi^2}\int_0^\infty\bigg\{\int_0^\infty\bigg[1-\frac{r_4\exp(r_2t)-r_3\exp(r_1t)}{r_2-r_1}\bigg]
\frac{\cos(y\xi)}{\xi^2}\,d\xi\bigg\}\times
\end{equation*}
\begin{equation}
\times\bigg\{\int_0^\infty\bigg[1-\frac{r_2r_3\exp(r_1t)-r_1r_4\exp(r_2t)}{r_2-r_1}\lambda\bigg]
\frac{\cos(y\xi)}{\xi^2}\,d\xi\bigg\}dy
\end{equation}
and
\begin{equation}
\delta=\int_0^\infty\bigg\{1-\frac{2}{\nu\pi t}\int_0^\infty\bigg[1-\frac{r_2r_3\exp(r_1t)-r_1r_4\exp(r_2t)}{r_2-r_1}\lambda\bigg]
\frac{\sin(y\xi)}{\xi^3}\,d\xi\bigg\}dy.                                              %18  \hskip2.1cm(3.3)
\end{equation}

By letting  $\lambda_r$ and $\lambda\rightarrow0$ or $\lambda\rightarrow\lambda_r$ into Eqs. (18)-(20), we attain to the similar expressions (cf. with Eqs. (3.4)-(3.6) from [6])
\begin{equation}
L_N=-\frac{2\rho l A^2t}{\pi}\int_0^\infty(1-\mbox{e}^{-\nu t\xi^2})
\frac{d\xi}{\xi^2},                                                    %19     \hskip7.5cm(3.4)
\end{equation}
\begin{equation}
\Phi_N=\frac{4\rho l A^2}{\nu\pi^2}\int_0^\infty
\bigg[\int_0^\infty(1-\mbox{e}^{-\nu t\xi^2})
\frac{\cos(y\xi)}{\xi^2}\,d\xi\bigg]^2dy                                    %20        \hskip5.35cm(3.5)
\end{equation}
and
\begin{equation}
\delta_N=\int_0^\infty\bigg[1-\frac{2}{\nu\pi t}\int_0^\infty
(1-\mbox{e}^{-\nu t\xi^2})
\frac{\sin(y\xi)}{\xi^3}\,d\xi\bigg]dy                                %21     \hskip5.6cm(3.6)
\end{equation}
corresponding to a Newtonian fluid performing the same motion. The similar expressions for Maxwell or second grade fluids are also obtained as limiting cases of (18)-(20) for  $\lambda_r$, respectively, $\lambda\rightarrow0$. The integrals in Eqs. (21)-(23) can be found and   $L_N$, $\Phi_N$ and $\delta_N$ can be written in simple forms
\begin{equation*}
L_N=-2\rho l A^2t\sqrt{\frac{\nu t}{\pi}},
\end{equation*}
\begin{equation*}
\Phi_N=\frac{8(\sqrt{2}-1)}{3}
\rho l A^2t\sqrt{\frac{\nu t}{\pi}},
\end{equation*}
and
\begin{equation}
\delta_N=\frac{4}{3}\sqrt{\frac{\nu t}{\pi}}.                                          %23        \hskip10.8cm(3.8)
\end{equation}
in which the boundary layer thickness is independent of the constant acceleration $A$. Indeed, using Eq. (A3) from appendix with $y=0$, Eq. $\mbox(24)_1$  is immediately obtained. In order to get the simplified form of $\Phi_N$, we must use Eqs. (A3), $\mbox{(A1)}_1$, (A4) and (A5). As regards the boundary layer thickness, from (23), (A6) and (A7) one obtains  $\mbox(24)_3$.
\section{Asymptotic approximations}
 \subsection{$\lambda/t$ and $\lambda/t\ll1$}
\indent\indent Let us now consider the case when the dimensionless relaxation and retardation  times  $\lambda/t$ and  $\lambda_r/t=\alpha/(\nu t)$ are much less then one. \emph{In this case the terms containing $\exp(r_2t)$  from Eqs. (9), (10) and (18)-(20) can be neglected, they going to zero faster then  $(\lambda/t)^2$.}  . Furthermore, for convenience (in order to get the results for second grade fluids as a limiting case) we shall use  $\alpha/(\nu t)$ instead of  $\lambda_r/t$. In these conditions, the next approximations
\begin{equation*}
\sqrt{(1+\alpha\xi^2)^2-4\nu\lambda\,\xi^2}=1+(\alpha-2\nu\lambda)\xi^2-2\nu\lambda(\alpha-\nu\lambda)\xi^4
+\cdot\cdot\cdot,
\end{equation*}
\begin{equation*}
\frac{1}{\sqrt{(1+\alpha\xi^2)^2-4\nu\lambda\xi^2}}=1-\alpha\xi^2+2\nu\lambda\xi^2+\cdot\cdot\cdot,
\end{equation*}
\begin{equation*}
\mbox{e}^{r_1t}=\mbox{e}^{-\nu t\xi^2}(1+\alpha\nu t\xi^4-\nu^2t\lambda\xi^4+\cdot\cdot\cdot);\,\,\,
\frac{r_3}{r_2-r_1}=-1+\alpha\xi^2-\nu\lambda\xi^2+\cdot\cdot\cdot,
\end{equation*}
\begin{equation}
\frac{r_3\mbox{e}^{r_1t}}{r_2-r_1}=-\mbox{e}^{-\nu t\xi^2}
[1+\alpha(\nu t\xi^2-1)+\nu \lambda\xi^2(1-\nu t\xi^2)+\cdot\cdot\cdot],                               %24         \hskip6.5cm(4.1)
\end{equation}
\begin{equation*}
\frac{\lambda r_2r_3\mbox{e}^{r_1t}}{r_2-r_1}=
\mbox{e}^{-\nu t\xi^2}(1-\nu^2 \lambda t\xi^4+\cdot\cdot\cdot),
\end{equation*}
are valid for each  $\xi$ and $t$ greater than zero.

Introducing $(24)_{5, 6}$  into (5) and (6) and having Eqs. (9) in mind, we find that
\begin{equation}
u(y,t)=u_N(y,t)-\frac{2\nu At}{\pi}\lambda
\int_0^\infty\xi\mbox{e}^{-\nu t\xi^2}\sin(y\xi)d\xi+\cdot\cdot\cdot                  %25         \hskip5.cm(4.2)
\end{equation}
and
\begin{equation}
\tau(y,t)=\tau_N(y,t)+\frac{2\mu A}{\pi}
\lambda\int_0^\infty(1-\nu t\xi^2)
\mbox{e}^{-\nu t\xi^2}\cos(y\xi)d\xi+\cdot\cdot\cdot.                         %26     \hskip4cm(4.3)
\end{equation}
Using now Eqs. $\mbox{(A.2)}_2$ , (A.8) and (A.9) from appendix, we can write the simpler forms
\begin{equation}
u(y,t)=u_N(y,t)-\frac{Ay}{2}\sqrt{\frac{t}{\nu\pi}}\exp
\bigg(-\frac{y^2}{4\nu t}\bigg)\cdot\frac{\lambda}{t}
+O\bigg[\bigg(\frac{\lambda}{t}\bigg)^2\bigg]                                 %27    \hskip3.7cm(4.4)
\end{equation}
and
\begin{equation}
\tau(y,t)=\tau_N(y,t)+\frac{\mu A}{2}\sqrt{\frac{t}{\nu\pi}}
\bigg(1+\frac{y^2}{2\nu t}\bigg)\exp\bigg(-\frac{y^2}{4\nu t}\bigg)\cdot\frac{\lambda}{t}
+O\bigg[\bigg(\frac{\lambda}{t}\bigg)^2\bigg].                                     %28         \hskip2.6cm(4.5)
\end{equation}
where
\begin{equation*}
\beta=Max\{\alpha/(\nu t), \lambda/t\}.
\end{equation*}
\section{Conclusion}
In this paper it is presented a study of the energetic balance corresponding to the unsteady flow of an Oldroyd-B fluid due to a constantly accelerating plate. Exact and approximative expressions are established for dissipation, the power due to the shear stress at the wall and the boundary layer thickness. As a consequence, the changing of the kinetic energy with time is obtained from the energy balance. In the special cases, when the relaxation or retardation time tends to zero, our results are going to the corresponding results for second grade and Maxwell fluids. If both times are going to zero, the similar results for Newtonian fluids are recovered.

Finally, by analogy with the Teipel's series expansion [7], series solutions have been established both for the velocity filed   and the adequate shear stress  . Our solutions, unlike the series solution of Teipel, are completely determined by means of the governing equations and the appropriate boundary conditions. Furthermore, as it was to be expected, these solutions are identical to those resulting from exact solutions by means of the asymptotic approximations.

In comparison or by analogy with the unsteady flow induced by a suddenly moved plate in Newtonian and non-Newtonian fluids, the next remarks can be made:

\begin{enumerate}
  \item The boundary layer thickness corresponding to these motions is independent of the constant velocity $U$ or the constant acceleration $A$ of the plate for all models.
  \item If  $\lambda>\lambda_r$, in comparison with Newtonian fluids, $L$ and $\phi$  increase for the flow induced by a suddenly moved plate and decrease for the flow produced by a constantly accelerating plate, while  $\delta$ decreases for both motions.
  \item The series solutions corresponding to the flow induced by a suddenly moved plate contain two free constants, although all conditions have been fulfilled, while the series solutions for the flow due to a constantly accelerating plate are completely determined by means of the initial and boundary conditions.
  \item The approximative solutions obtained by means of the series expansions are identical to those resulting from the exact solutions using asymptotic approximations.
\end{enumerate}
\section*{Acknowledgement}

The authors acknowledge support from the Ministry of Education and Research, through PN II-program, Contract: PN II-2007.

{}


\begin{thebibliography}{00}
\bibitem{1} K. Bühler, J. Zierep, Energetische Betrachtungen zum Rayleigh - Stokes problem, Proc. Appl. Math. Mech. PAMM 5, 539-540 (2005).
\bibitem{1} J. Zierep, C. Fetecau, Energetic balance for the Rayleigh - Stokes problem of a second grade fluid, Int. J. Eng. Sci. 45, 155-162 (2007).
\bibitem{1} J. Zierep, C. Fetecau, Energetic balance for the Rayleigh - Stokes problem of a Maxwell fluid, Int. J. Eng. Sci. 45, 617-627 (2007).
\bibitem{1} C. Fetecau, T. Hayat, J. Zierep, M. Sajid, Energetic balance for the Rayleigh-Stokes problem of an Oldroyd-B fluid, sent for publication.
\bibitem{1} C. Fetecau, D. Vieru, Corina Fetecau, M. Khan, Energetic balance for the flow induced by a constantly accelerating plate in a second grade fluid, sent for publication to Applied Mathematical Modeling.
\bibitem{1} Corina Fetecau, D. Vieru, A. Mahmood, C.Fetecau, On the energetic balance for the flow of a Maxwell fluid due to a constantly accelerating plate, sent for publication to Acta Mechanica.
\bibitem{1} I. Teipel, The impulsive motion of a flat plate in a visco - elastic fluid, Acta Mech. 39 277 - 279 (1981).
\bibitem{1} K.R. Rajagopal, A.R. Srinivasa, A thermodynamic frame work for rate type fluid models, J. Non-Newton. Fluid Mech. 88, 207-227 (2000).
\bibitem{1} R.B. Bird, R.C. Armstrong, O. Hassager, Dynamics of polymer liquids, Vol I, Fluid Mechanics, New York, Wiley, 1987.
\bibitem{1} J.G. Oldroyd, On the formulation of rheological equations of state, Proc. R. Soc. London Ser. A 200, 523-541 (1950).
\bibitem{1} C. Fetecau, Sharat C. Prasad, K. R. Rajagopal, A note on the flow induced by a constantly accelerating plate in an Oldroyd-B fluid, Appl. Math. Model. 31, 647 - 654 (2007).
\bibitem{1} C. Fetecau, J. Zierep, On a class of exact solutions of the equations of motion of a second grade fluid, Acta Mech. 150, 135 - 138 (2001).
\bibitem{1} M. E. Erdogan, On unsteady motions of a second-order fluid over a plane wall, Int. J. Non-Linear Mech. 38, 1045-1051 (2003).
\bibitem{1} D.D. Joseph, Fluid Dynamics of viscoelastic liquids, New York, Springer Verlag 1990.
\bibitem{1} D. Vieru, Corina Fetecau, C. Fetecau, Exact solutions for the flow of an Oldroyd-B fluid due to an infinite flat plate, to appear in ZAMP, 2007.
\end{thebibliography}
\end{document}